\documentclass[aps,amsfonts,pre,twocolumn,superscriptaddress,showpacs]{revtex4}
\usepackage{epsfig,amsmath,amssymb,bm,epsf,graphics,psfrag,verbatim,subfigure}

\def\jpos{\mathbf{r}} 
\def\jor{\hat{n}} 
\def\weight{\Omega} 
\def\patchsize{\phi_0} 
\def\kt{\kappa_t} 
\def\kb{\kappa_b} 
\def\jdisp{\mathbf{u}} 
\def\bendvec{\hat{b}} 
\def\twistvec{\hat{t}} 

\begin{document}

\title{Self-assembly of three-dimensional open structures using patchy colloidal particles}

\author{D. Zeb Rocklin}
\affiliation{Department of Physics, University of Michigan,
450 Church Street, Ann Arbor, MI 48109, USA}

\author{Xiaoming Mao}
\affiliation{Department of Physics, University of Michigan,
450 Church Street, Ann Arbor, MI 48109, USA}

\date{\today}

\begin{abstract}
Open structures can display a number of unusual properties, including a negative Poisson's ratio, negative thermal expansion, and holographic elasticity, and have many interesting applications in engineering.  However, it is a grand challenge to self-assemble open structures at the colloidal scale, where short-range interactions and low coordination number can leave them mechanically unstable. 
In this paper we discuss the self-assembly of open structures using triblock Janus particles, which have two large attractive patches that can form multiple bonds, separated by a band with purely hard-sphere repulsion. Such surface patterning leads to open structures that are stabilized by orientational entropy (in an “order-by-disorder” effect) and selected over close-packed structures by vibrational entropy.  For different patch sizes the particles can form into either tetrahedral or octahedral structural motifs which then compose open lattices, including the pyrochlore, the hexagonal tetrastack and the perovskite lattices.  Using an analytic theory, we examine the phase diagrams of these possible open and close-packed structures for triblock Janus particles and characterize the mechanical properties of these structures.  Our theory leads to rational designs of particles for the self-assembly of three-dimensional colloidal structures that are possible using current experimental techniques.
\end{abstract}
\maketitle

\section{Introduction}
\label{sec:intro}
Open structures are those in which the microscopic constituents occupy only a low fraction of the total volume, leaving open space between them and allowing the structures to undergo a richer variety of fluctuations and deformations than close-packed structures.  They possess striking properties such as negative Poisson's ratio~\cite{Greaves2011,Sun2012,Kapko2009,Grima2007}, negative thermal expansion~\cite{Ernst1998,Hammonds1996}, and beyond~\cite{Davis1992, McDowellboyer1986}, leading to many interesting applications in engineering.

A structure can be open at different scales, depending on the size of the constituents and the open space.  Zeolite, a natural aluminosilicate mineral, is an example of an open lattice structure with pores (open spaces) at the scale of Angstroms~\cite{Breck1973,Davis1992}.  Nature also offers us structures that are open at larger length scales, such as foams and bones, but these structures are generally disordered.  

\begin{widetext}

\begin{figure}[hh]
\centering
\includegraphics[width=.9\textwidth]{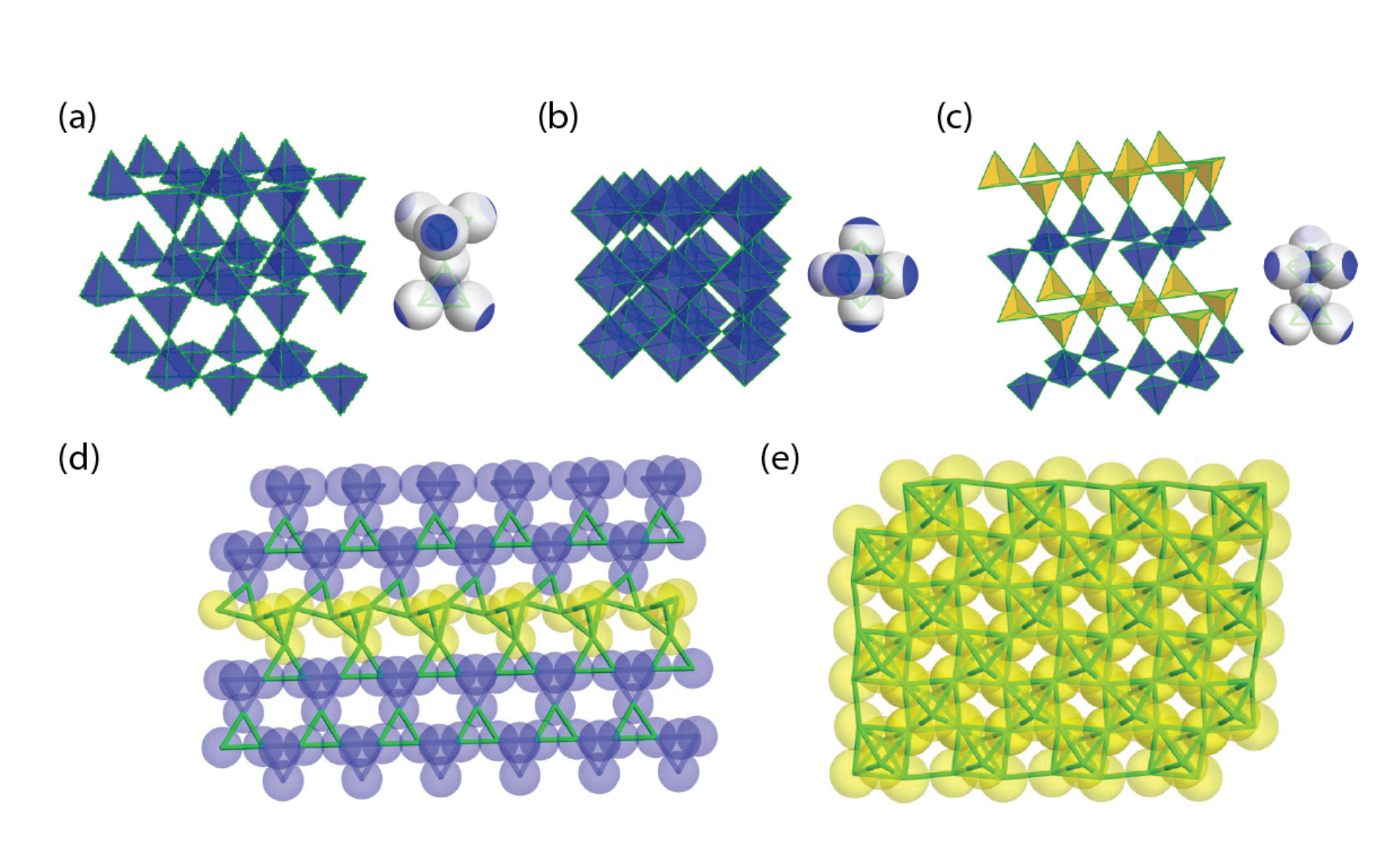}
\caption{Open lattices (a-c, where structures of the lattices are shown on the left and the triblock Janus particles forming the basic motifs are shown on the right) and the floppy modes (d, e) such structures have if composed of isotropic particles rather than triblock Janus particles.  
(a) Pyrochlore lattice, with tetrahedral cells and six bonds for each triblock Janus particle. (b) Perovskite lattice, with octahedral cells and eight bonds per particle. (c) Hexagonal tetrastack (HT) lattice. Otherwise identical to the perovskite structure, blue and yellow layers are rotated $\alpha=60^\circ$ relative to one another, so that the bonds of the triblock Janus particles that lie at the juncture between layers have a permanent twist.
(d) Floppy modes in a layer of the pyrochlore lattice. The yellow row of tetrahedral cells has been rotated showing one floppy mode which is zero energy if the composing particles are isotropic.  (e) A layer of the perovskite lattice. Zero energy modes rotate all of the octahedral cells in a layer relative to their neighbors.  These floppy modes can lead to the collapse of open lattices if composing particles are isotropic with short-range interactions.
}
\label{fig:latticediagrams}
\end{figure}

\end{widetext}

Obtaining open structures with pores at the colloidal scale is not only desirable for such applications as photonic crystals~\cite{Joannopoulos1997,Moroz2002,Garcia-Adeva2006,Galisteo-Lopez2011}, but also presents a fundamentally interesting question in physics.  The challenge comes from the proximity of these open structures to \emph{mechanical instability}, which makes them floppy against collapsing into more close-packed structures.  The stability of a colloidal structure can be explained by the counting argument due to Maxwell~\cite{Maxwell1864}: that for a structure to be mechanically stable, its total number of constraints must equal or exceed its total number of internal degrees of freedom.  This argument, applied to simple colloidal particles with only central force interactions between nearest neighbor particles, leads to the stability criterion $z>z_c=2d$ where $z$ is the coordination number and $d$ is the spatial dimension, and the special state with $z=z_c$ is called the \emph{isostatic point}.  According to this criterion, most open lattices are not stable.  For example, the diamond lattice has $z=4<2d$ and is below isostaticity and the pyrochlore lattice has $z=6=2d$ and is at the verge of instability.  Another interesting example is the perovskite lattice, which has $z=8>2d$ but still has some floppy modes due to redundancy of constraints.  Fig.~\ref{fig:latticediagrams} shows examples of these lattices and their floppy modes.

Thus, to obtain open lattice structures at the colloidal scale additional interactions beyond a simple nearest neighbor isotropic potential have to be added to provide mechanical stability, preventing the structure from collapsing.  Various designs have been proposed~\cite{Cohn2009,Edlund2011,Torquato2009,Velikov2002,Glotzer2007,Romano2012,Tkachenko2002}, which require complicated inter-particle potentials that are difficult to realize in experiment.  

Remarkably, an open two-dimensional kagome lattice has been self-assembled using triblock Janus colloidal particles very recently~\cite{Chen2011}.  Janus colloidal particles have chemical coatings that cover a fraction of their surfaces (\lq\lq patches\rq\rq), making their interaction potential anisotropic.  The triblock Janus particles used in the experiment in Ref.~\cite{Chen2011} are characterized by patches at their north and south poles of short range hydrophobic attraction of depth $\sim 10 k_B T$ and a middle band of strongly screened electrostatic repulsion~\cite{Jiang2010,Chen2011b}.  The short range nature of the inter-particle potential in this triblock Janus particle system does not directly provide more energetic constraints for stability.  Instead, it has been found that entropy plays the important role of stabilizing open lattices formed by these triblock Janus particles because, among all structures of degenerate potential energy, the open lattice permits the most rotational and vibrational fluctuations of the constituent particles~\cite{Mao2013,Mao2013b}.

In this work, we explore the self-assembly of three-dimensional lattices using the triblock Janus particle based on this entropic stabilization/selection mechanism.  We show that depending on the size of the attractive patches, these particles can either form a perovskite lattice or a mixed phase of pyrochlore/hexagonal tetrastack (HT) lattices (Fig.~\ref{fig:latticediagrams}) and that under pressure they can collapse into a close-packed face centered cubic (FCC) lattice.  
The self-assembly of the pyrochlore/HT lattices has been examined in Ref.~\cite{Romano2012} using simulations.  In this work we analytically describe the role of entropy in selecting the phases and characterize the phase diagram of this system using a harmonic approximation of the entropic effects and lattice dynamics calculations.  Our theory opens the door to greatly simplified designs of building blocks for the self-assembly of three-dimensional open lattices.

\section{Model}
\subsection{Statistics of patchy particles}
\label{sec:stat}

Generally, we may describe the equilibrium statistical mechanics of a set of anisotropic particles via the partition function~\cite{Mao2013,Mao2013b}
\begin{eqnarray}
\label{eq:partitionfunction}
\mathcal{Z}= \int \exp \left[-\frac{1}{k_B T} H\left( \{\jpos_j,\jor_j\}\right)\right]
\prod_j d \jpos_j d \jor_j,
\end{eqnarray}
\noindent where the Hamiltonian $H$ depends on the positions $\jpos_j$ and $\jor_j$ orientations of the particles and $k_B$ is the Boltzmann constant and $T$ the temperature.

\begin{figure}[hh]
\centering
\includegraphics[width=.35\textwidth]{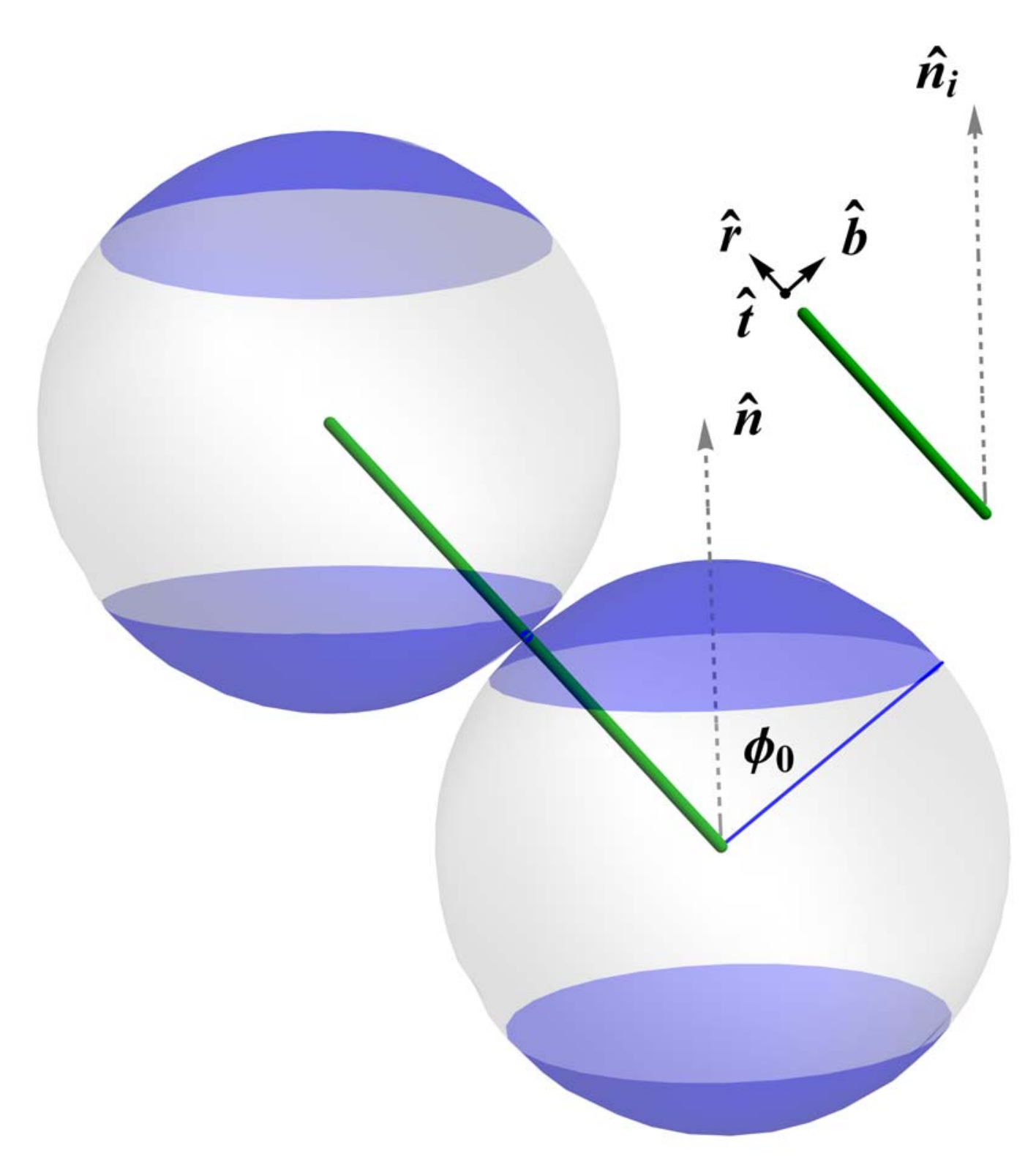}
\caption{
Triblock Janus particles have two circular attractive patches separated by a repulsive band. The orientation of a triblock Janus particle $i$ is given by its ``north pole'', $\hat{n}_i$. The patch size, $\phi_0$, is the angle between the edge of a patch and its center. Two adjacent particles are bonded attractively if the contact point lies within contact patches on both particles. The movement of a particle relative to particle $i$ can be decomposed into the radial direction, $\hat{r}$, the bending direction $\bendvec$ which is orthogonal to $\hat{r}$ but points toward the preferred axis of particle $i$, and the twisting direction $\twistvec$ which is orthogonal to both $\hat{r}$ and the preferred $\hat{n}_i$.
}
\label{fig:tjp}
\end{figure}

The particles we consider, \emph{triblock Janus particles}, have two attractive patches separated by a repulsive band, as shown in Fig.~\ref{fig:tjp}. These particles have a hardcore repulsion when their separation is close to their diameter, $a$. Additionally, there is a short-range attraction between particles whose patches are oriented towards one another, permitting the formation of bonds.

We now consider a lattice of such particles, which may in general include both attractive bonds, where particles contact each other in their attractive patches, and repulsive bonds, where the particles are held close to each other by the lattice structure but derive no energetic benefit. The strength of the attractive bonds can be around $\sim 10 k_B T$~\cite{Jiang2010,Chen2011b}, so thermal fluctuations will not give configurations with broken bonds appreciable Boltzmann weight.
We may then restrict our ensemble to a particular bond structure and integrate out the particle orientations $\jor_j$ from the partition function of Eq.~(\ref{eq:partitionfunction}), and the result will be a degeneracy factor $\weight_j(\{\jpos_i\})$ proportionate to the number of orientations (all of which share the same potential) which keep all the attractive bonds within attractive patches. The orientational entropy of a particle depends on the relative positions (but not orientations, given that in a lattice we take any configurations breaking an attractive bond to have Boltzmann factor zero) of each particle with which it has an attractive bond, and is related to the degeneracy factor via~\cite{Mao2013,Mao2013b}
\begin{eqnarray}
s_j = k_B \ln \weight_j \left(\{\jpos_i\}\right).
\end{eqnarray}
\noindent This term accounts entirely for the effect of particle orientations given in Eq.~(\ref{eq:partitionfunction}), leading to an effective Hamiltonian which depends only on particle \emph{positions}.


For the case of a triblock Janus particle $j$ with unit vectors $\hat{e}_{ij}$ to each of its attractively-bonded neighbors $i$. These bonds lie within an attractive patch provided that
\begin{eqnarray}
\left| \hat{e}_{ij} \cdot \jor_j  \right| \ge \cos \patchsize,
\end{eqnarray}
\noindent where $\patchsize$ is the patch size, the maximum angle between a point in the patch and the particle's ``north pole'', the center of an attractive patch. Positive (negative) values of $\hat{e}_{ij} \cdot \jor_j$ correspond to bonds that lie in the particle's northern (southern) hemisphere.
Thus, the degeneracy factor is simply
\begin{eqnarray}
\label{eq:omegaj}
\weight_j(\{\jpos_i\}) = \int d \jor_j \prod_i \Theta \big[ \left|\hat{e}_{ij} \cdot \jor_j  \right| - \cos \patchsize \big],
\end{eqnarray}
\noindent where $\Theta(\cdot)$ is the Heaviside step function.

This entropy is maximized when all bonds lie as close as possible to either the north or south pole of a particle as is permitted by the hardcore repulsion between particles.  Three-dimensional open lattices can then be formed from
 rigid octahedral or tetrahedral cells (depending on whether the patches are large enough to support four bonds or only three) that meet at their corners (centers of triblock Janus particles) and directly oppose one another there.
This permits the formation of three types of three-dimensional lattices: the pyrochlore, the Hexagonal Tetrastack (HT) and the perovskite, cf. Fig.~\ref{fig:latticediagrams}. 

In the pyrochlore and HT lattices, a particle forms a rigid tetrahedral structure with its three northern neighbors, and another with its three southern neighbors. In the perovskite, any particle lies at a vertex joining two octahedra and has four bonds in each of its attractive patches.  The cell structure is determined by the patch size: above $\phi_0^{\textrm{min}(4)}=45^\circ$ four bonds per patch become possible (and energetically favored), leading to octahedral cells. Between this patch size and $\phi_0^{\textrm{min}(3)} = \arccos(\sqrt{2/3}) \approx 35.3^\circ$ only three bonds are possible, leading to tetrahedral cells.

Any distortions of these rigid polyhedra will carry finite energy costs but, as discussed in the introduction, floppy modes exist in these lattices in which only the \emph{angles} between neighboring polyhedra change (Fig.~\ref{fig:latticediagrams}).  Floppy modes in these lattices have been named Rigid Unit Modes (RUMs)~\cite{Hammonds1996}.  Such RUMs are the dominant thermal fluctuation present in a lattice of triblock Janus particles, with the bond length fluctuations much smaller than bond angle fluctuations~\cite{Mao2013}.

\subsection{Analytic model of bending rigidity}

\begin{figure}[hh]
\centering
\includegraphics[width=.3\textwidth]{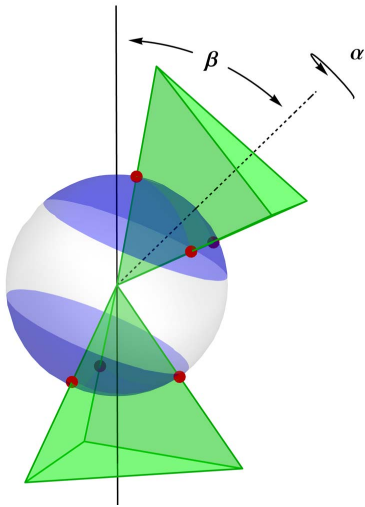}
\caption{
The tetrahedral cell that a triblock Janus particle forms with its neighbors in one patch may be twisted through an angle $\alpha$ about its axis and bent through an angle $\beta$ relative to the other tetrahedral cell. As the bending angle increases, fewer orientations of the central triblock Janus particle are possible, lowering the weight of such configurations in the lattice and creating an effective modulus against bending.
}
\label{fig:alphabetadiagram}
\end{figure}

In order to analyze the dynamics of the three-dimensional lattice, we approximate the effects of a particle's orientational entropy through an analytic effective energy incorporating only the rigid body rotations of a particle's northern bonds about its southern ones:
\begin{eqnarray}
\label{eq:harmapprox}
U_r = \frac{1}{2} \kt \alpha^2 + \frac{1}{2} \kb \beta^2.
\end{eqnarray}
\noindent The \textit{bending angle} $\beta$, which can be decomposed into two components, $\beta_x$ and $\beta_y$, is the angle by which the northern cell differs from perfectly opposing the southern cell, as depicted in Fig.~\ref{fig:alphabetadiagram}. The \textit{twisting angle} $\alpha$ is the angle by which the northern cell is rotated about its own axis, with $\alpha=0$ corresponding to northern bonds that are (for $\beta=0$) collinear with southern ones.

In the harmonic approximation of Eq.~(\ref{eq:harmapprox}), the mean square bending of a free triblock Janus particle---that is, one whose bonds are free to undergo any rigid rotation---is related to the bending modulus via

\begin{eqnarray}
\langle \beta^2 \rangle = 2 \, \frac{k_B T}{\kb}.
\end{eqnarray}

\begin{figure}[hh]
\centering
\includegraphics[width=.45\textwidth]{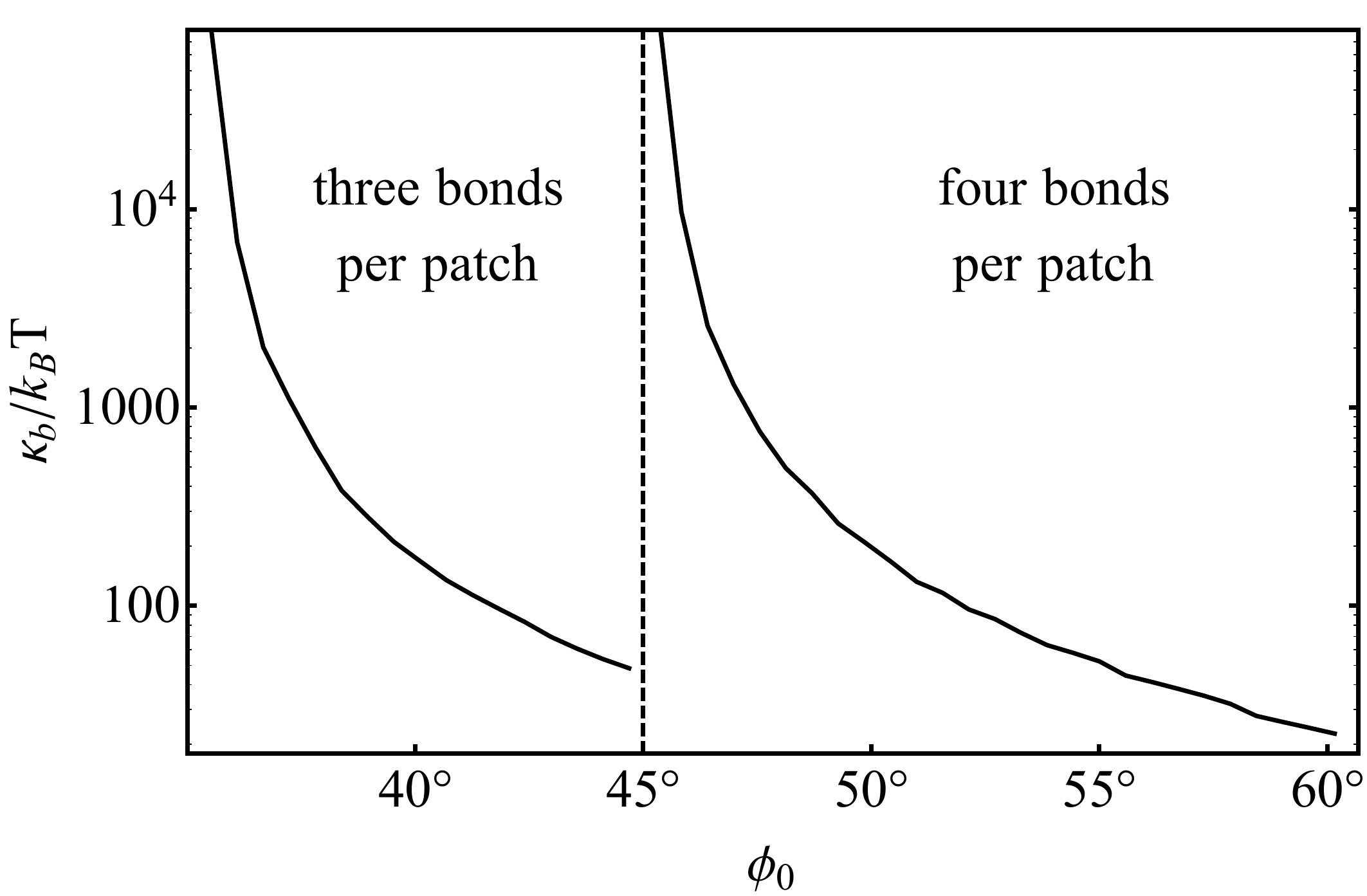}

\caption{
The effective bending modulus $\kb$, in units of temperature, as a function of $\phi_0$, the angular half-width of the attractive patch. Between $\phi_0 \approx 35.3^\circ$ and $\phi_0=45^\circ$ the patch is able to support a maximum of three bonds, as in the pyrochlore lattice. Above $45^\circ$, the patch supports four bonds, as in the perovskite lattice.
}
\label{fig:kb}
\label{fig:patchsizevsmodulus}
\end{figure}

\noindent We can determine $\langle \beta^2 \rangle$ numerically via Eq.~(\ref{eq:omegaj}) and thereby obtain the effective bending modulus for any patch size by requiring that it reproduce this first moment. For patch sizes very close to the minimum sizes necessary to support all of the bonds, the bending modulus diverges as $\left(\phi_0 - \phi_0^{\textrm{min}}\right)^{-2}$, as depicted in Fig.~\ref{fig:patchsizevsmodulus}. Within our analytic model, it is this bending modulus that imposes an effective cost on otherwise zero-energy modes and ensures that bending and twisting angles remain small.

For a free triblock Janus particle, the twisting modulus $\kt$ proves to be zero.  This is because for any twisting angle $\alpha$, the integral of the weight $\Omega$ over different bending angles leads to the same value. 

This does not mean, however, that large twisting angles occur in a lattice.  Bending or twisting the bonds of a given triblock Janus particle can only be accomplished by bending and twisting other bonds in the lattice. Thus, certain values of $\alpha$ are not accessible in a lattice because they necessarily involve bending angles that break bonds (e.g., for a patch size that requires $\beta \le 1^\circ$, $\alpha = 20^\circ$ is not accessible). Thus, in a lattice the twisting angles will be restricted to small deviations from their optimum value (which is $\alpha=0^\circ$ except for in particles at junctions between layers in the HT lattice, for which it is $\alpha = 60^\circ$), and an effective twisting modulus up to $\sim k_B T$ arises~\cite{Mao2013}.  As we will see below, the bending modulus gives all the zero modes of the lattice an effective energy, and the inclusion of a finite twisting modulus in our model would have only a very small effect.

\begin{figure}[hh]
\centering
\subfigure[]{\includegraphics[width=.23\textwidth]{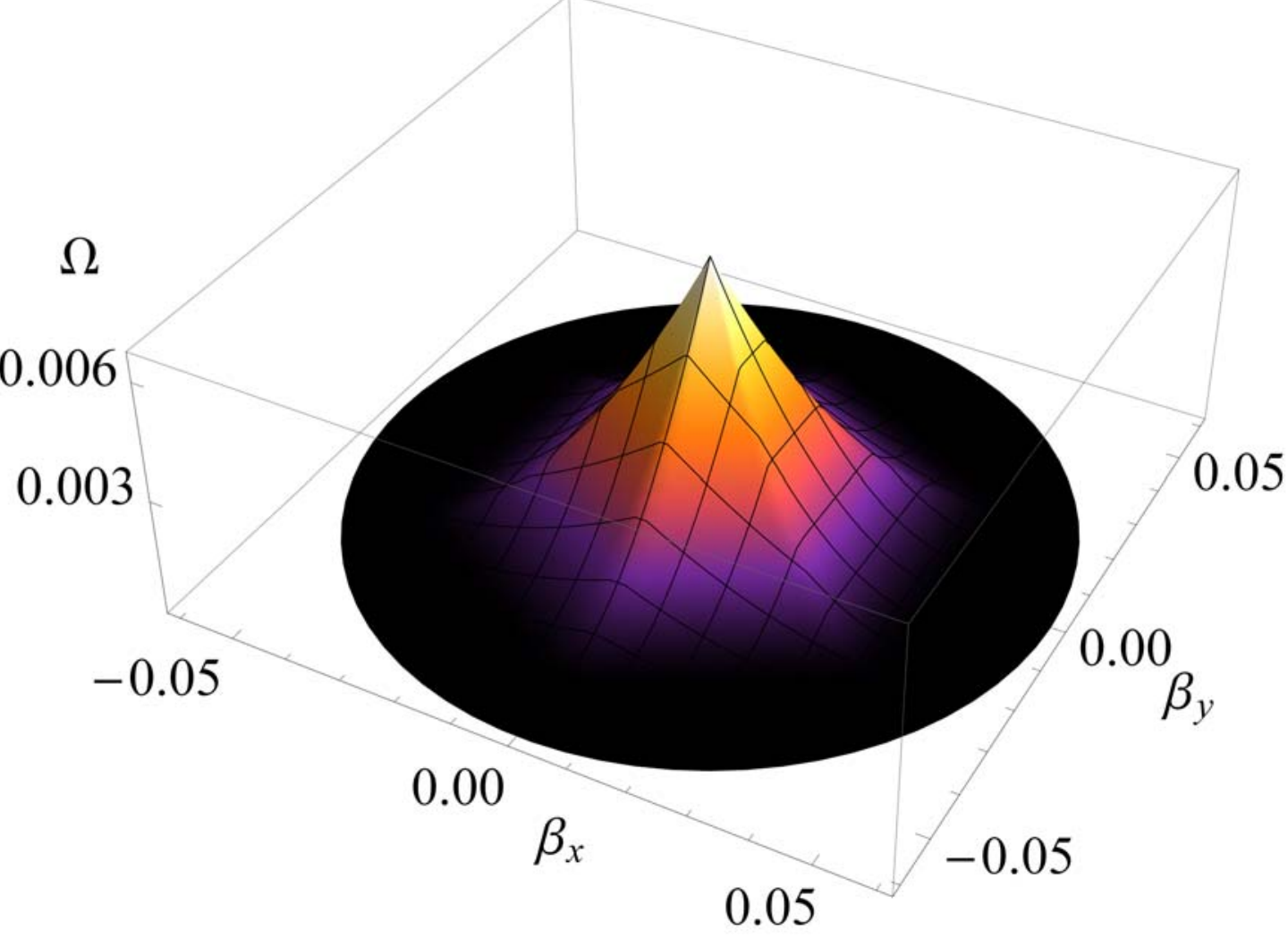}}
\subfigure[]{\includegraphics[width=.23\textwidth]{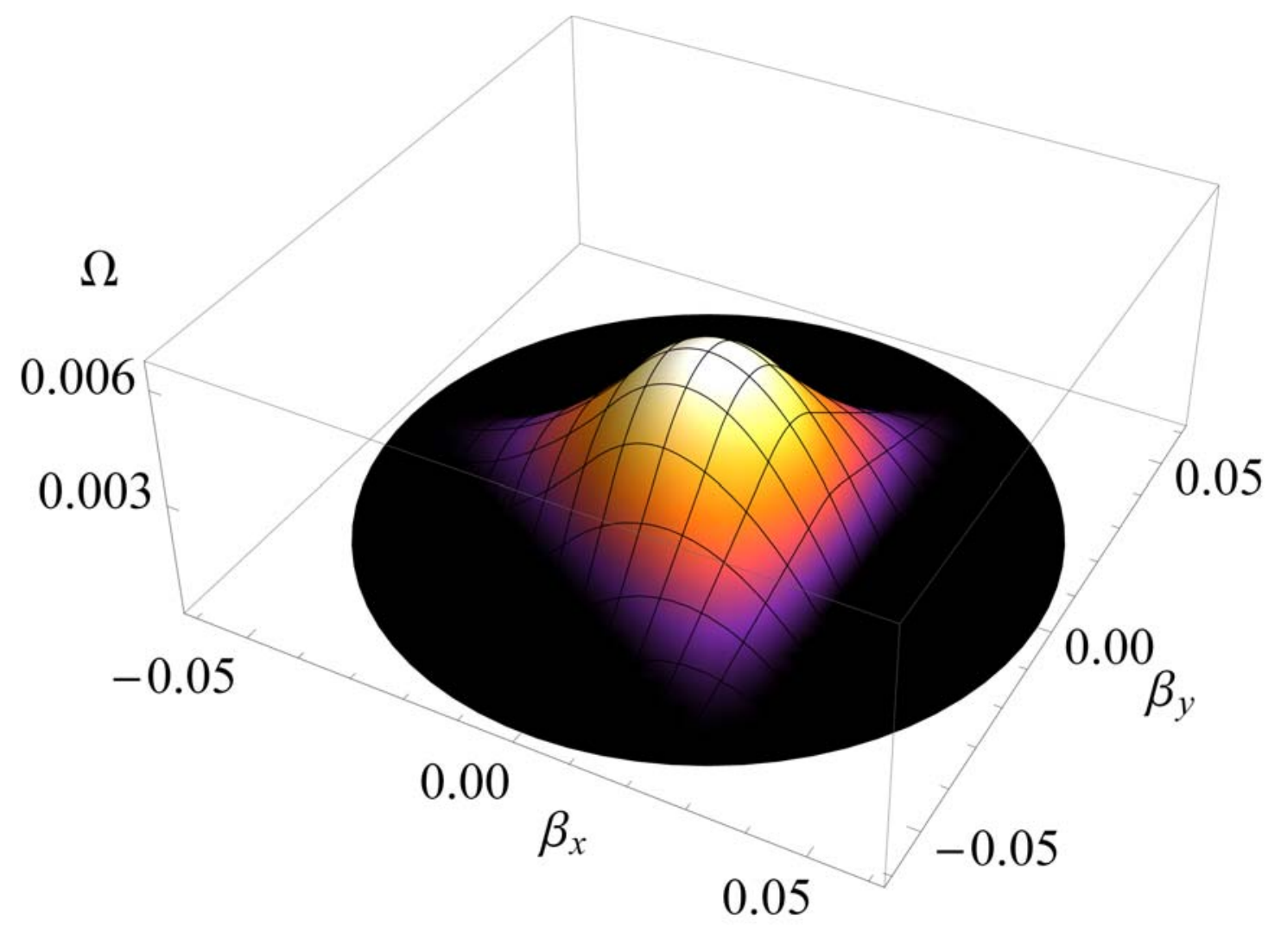}}

\subfigure[]{\includegraphics[width=.23\textwidth]{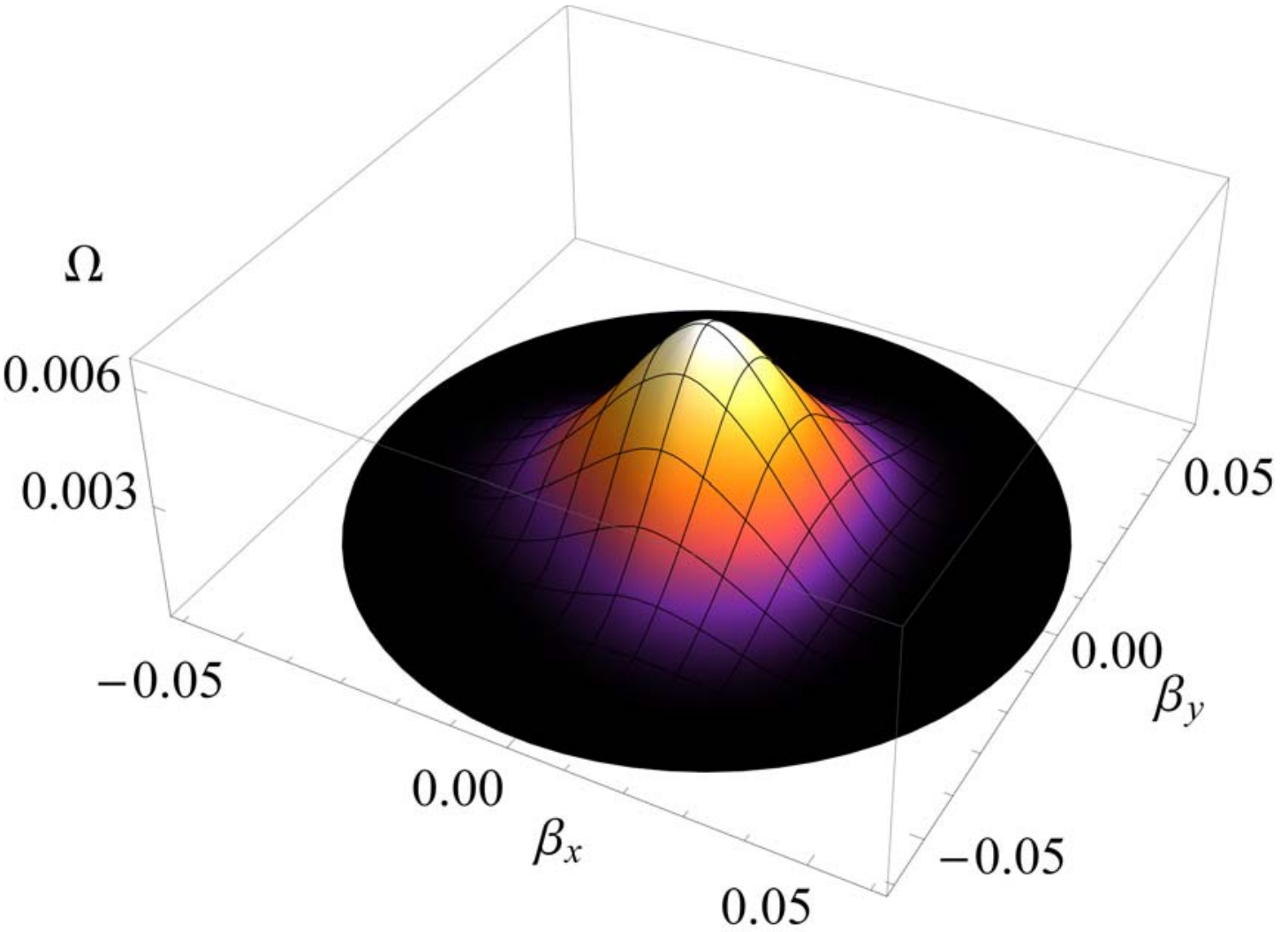}}
\subfigure[]{\includegraphics[width=.23\textwidth]{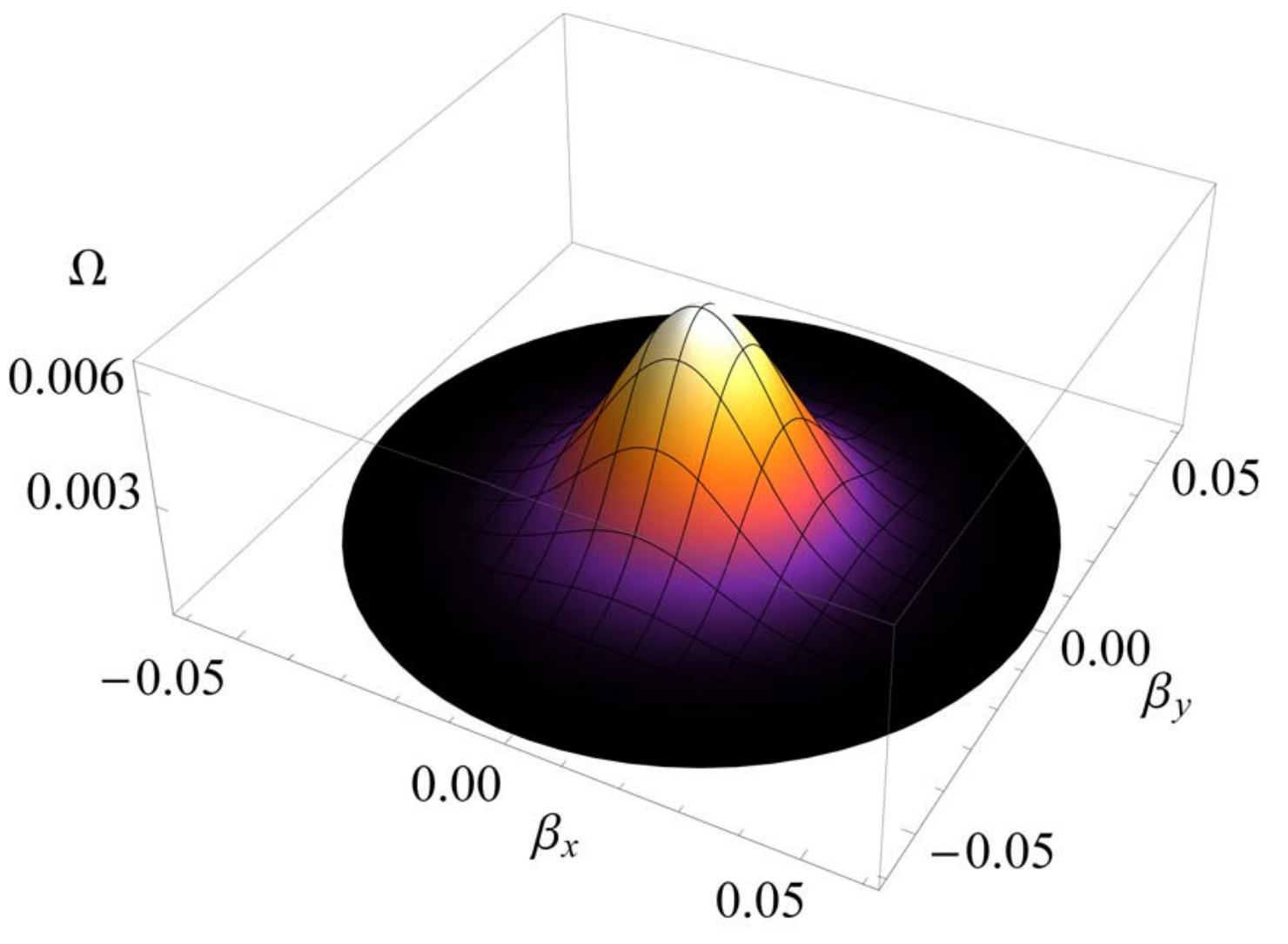}}

\caption{
The weight $\Omega$, or total fraction of triblock Janus particle orientations associated with bending angle components $(\beta_x,\beta_y)$ of a triblock Janus particle with patch size $\phi_0 = 36.1^\circ$ barely large enough to support its three bonds per patch. In (a), twisting angle $\alpha$ is fixed at $0^\circ$, as in the pyrochlore lattice, and there is hexagonal symmetry associated with the six bond directions. In (b), $\alpha = 60^\circ$, as in the HT structure, leading to triangular symmetry as some bonds oppose one another. In (c), all values of $\alpha$ are averaged over, as for isolated tetrahedral cells not held in place by the lattice. In (d), the harmonic approximation to the bending rigidity used in the analytic model is rotationally symmetric.
}
\label{fig:ohms}
\end{figure}

The dependence of $\Omega_j$ on the bending angle in each of these scenarios is shown in Fig.~\ref{fig:ohms}.

\subsection{Lattice energetics}
\label{sec:energetics}

Having accounted for the effects of orientational entropy, we may now construct an effective Hamiltonian for a lattice purely in terms of the \textit{positions} of the triblock Janus particles. In the lattice, we may take these positions as
\begin{eqnarray}
\jpos_j = \mathbf{R}_j+\jdisp_j,
\end{eqnarray}
\noindent where $\mathbf{R}_j$ is the lattice site and the displacement $\jdisp_j$ is small compared to the particle spacing. This then allows us to formulate the central force interaction between bonded particles as a harmonic spring, so that
\begin{eqnarray}
V_{\textrm{cf}}(\jdisp_i-\jdisp_j) = \frac{k_{a(r)}}{2}\frac{\left[\left(\mathbf{R}_i-\mathbf{R}_j\right)\cdot \left(\jdisp_i - \jdisp_j\right)\right]^2}{\left(\mathbf{R}_i-\mathbf{R}_j\right)^2},
\end{eqnarray}
\noindent where the effective spring constant is $k_{a(r)}$ for attractive (repulsive) bonds. Thus, the effective Hamiltonian, incorporating both the central-force and rotational-entropy terms, is
\begin{eqnarray}
H_{\textrm{eff}} = \sum_{\langle i,j \rangle} V_{\textrm{cf}}(\jdisp_i-\jdisp_j) +
\sum_i \frac{\kb}{2} \beta_i^2,
\end{eqnarray}
\noindent where the first sum is over bonded neighbors, with $k_{i,j}$ differing depending on whether the bond is attractive or repulsive. 

General lattice displacements will distort as well as rotate the tetrahedral cells depicted in Fig.~\ref{fig:tjp}. However, the angle by which the bending components of the displacements--- combined vectorially--- rotates the tetrahedron from its $\jdisp=0$ position remains well-defined. The bending angle $\beta_i$ is the change in the angle between the surface normal vectors of the two tetrahedra (or octahedra) in which particle $i$ lies. Note that this depends on the positions of all of particle $i$'s neighbors, leading to an effective next-nearest neighbor coupling.

\begin{figure}[hh]
\centering
\subfigure[]{\includegraphics[width=.4\textwidth]{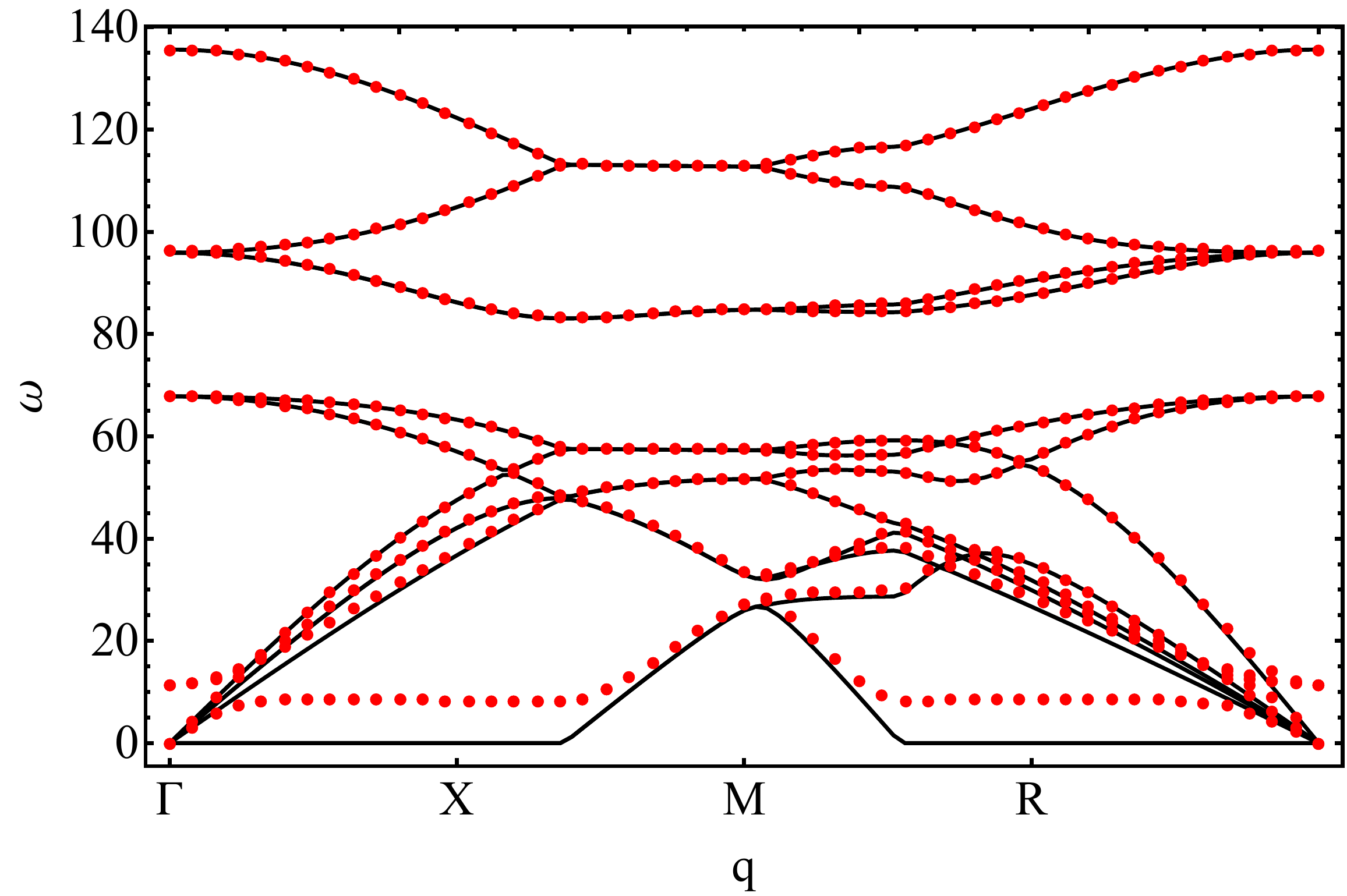}}

\subfigure[]{\includegraphics[width=.4\textwidth]{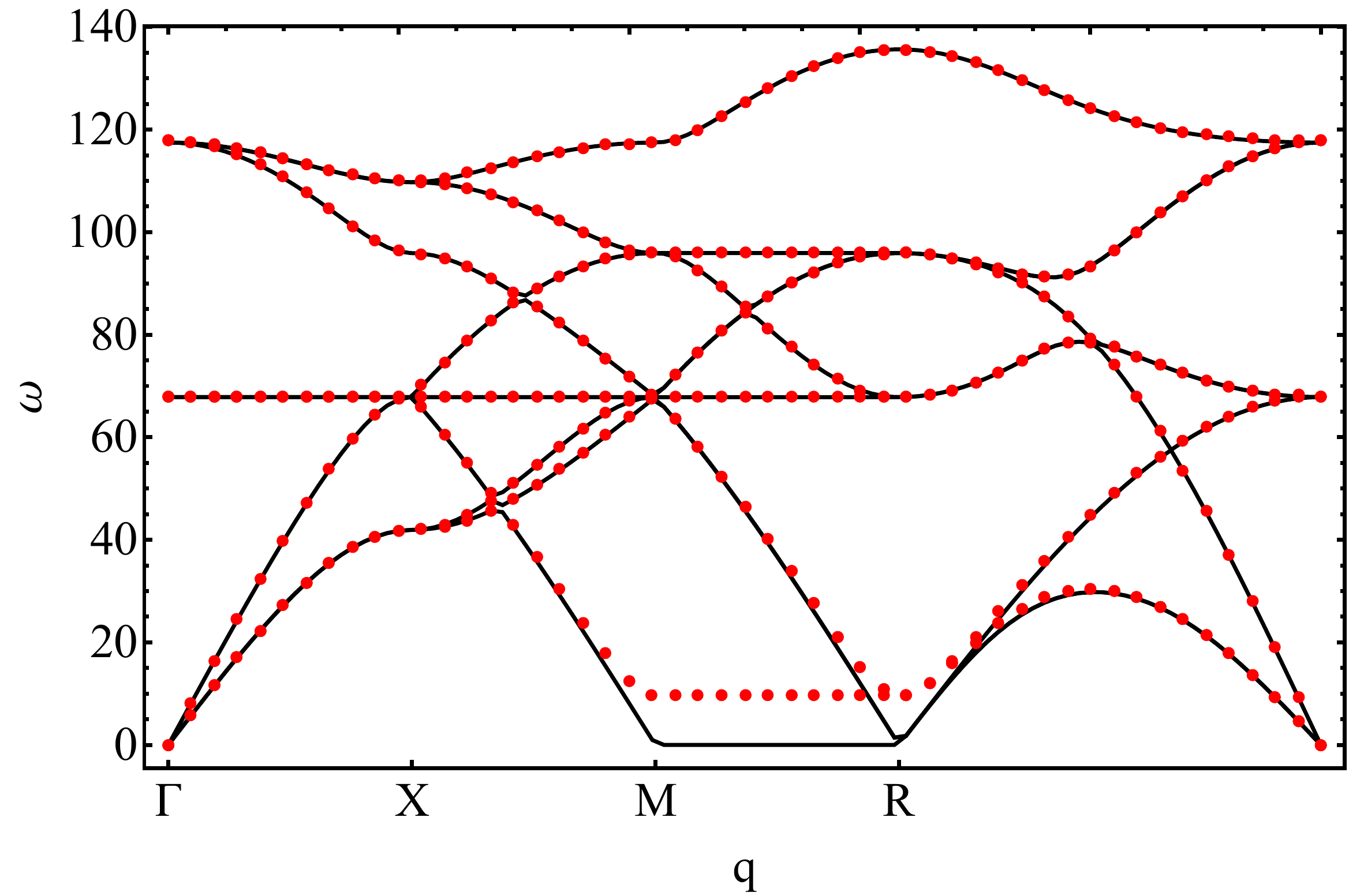}}

\caption{
Phonon dispersions for the pyrochlore (a) and perovskite (b) lattices. The black solid curves show the case of $k_a a^2= 2300 k_B T$ and $\kb = 0$, which includes a plane of zero modes in the pyrochlore lattice and a line of zero modes in the perovskite lattice, agreeing with the counting of floppy modes discussed above. These modes are lifted by the inclusion of $\kb = 33 k_B T$ (red dotted curves).  The values of $k_a$ and $\kb$ are taken from Ref.~\cite{Mao2013}.
}
\label{fig:dispersions}
\end{figure}

From this harmonic Hamiltonian, the dynamical matrix associated with a crystal lattice may be generated using standard techniques~\cite{Ashcroft1976}. The phonon dispersions for the dynamical matrices associated with the pyrochlore and perovskite lattices are shown in Fig.~\ref{fig:dispersions}.

In a pyrochlore lattice with $L^3$ tetrahedral cells the high-symmetry planes have the structure of the kagome lattice. As discussed in~\cite{Mao2013, Mao2013b}, each such kagome lattice has $O(L)$ zero modes consisting of twists applied to lines of triangular cells. The pyrochlore lattice therefore has $O(L^2)$ zero-energy (for $\kb=0$) modes in which lines of tetrahedra rotate. The HT lattice similarly has $O(L^2)$ zero modes, with $1/3$ of the kagome modes replaced by displacements of columns of tetrahedra in the direction perpendicular to the blue and yellow planes in Fig.~\ref{fig:latticediagrams}. In the perovskite lattice, which is above the isostatic point, the only zero modes are a number $O(L)$ in which an entire \emph{plane} of octahedra rotate in concert. These zero modes show up as zero frequency lines in Figure~\ref{fig:dispersions}.

For the self-assembly examined in~\cite{Mao2013}, the bending modulus is large in the sense that $\kb \approx 33 k_B T$ permits only small bending angles, but small compared to the central forces modulus, $k_a a^2 \approx 2300 k_B T$. As such, the effect of orientational entropy amounts only to a linear perturbation in the energy values associated with most of the modes of the lattice. However, for the zero modes, there is a $O(\sqrt{\kb})$ contribution to the increase in energy, as seen in Figure~\ref{fig:dispersions}. The effective bending modulus has a similar effect on the zero modes of the perovskite lattice.

In the absence of this orientational entropy effect, the zero modes of the lattices would permit large distortions of the lattices, destroying mechanical stability. The patchiness of the particles prevents arbitrary rotations of the tetrahedral units, and the orientational entropy of the triblock Janus particles favors the $\beta=0$ configuration over distortions to the lattices.

At finite temperature and fixed volume, the phase of the particles is determined by their free energy. As discussed in~\cite{Mao2013b}, the free energy per particle (not including a constant bonding energy) may be expressed in terms of the momentum-space dynamical matrix $\mathcal{D}(\vec{q})$ as
\begin{eqnarray}
f(\kb)= \frac{k_B T}{2 n_c} v_0 \int_{\mbox 1 BZ} \frac{d^d \vec{q}}{\left(2 \pi\right)^d } \ln \det \mathcal{D}(\vec{q},\kb),
\end{eqnarray}
\noindent where $v_0$ is the volume of a unit cell, and $n_c$ is the number of particles per unit cell.

\section{Results and Discussion}
\label{sec:discussion}

\subsection{Dependence of free energy on patch size}
\begin{figure}[hh]
\centering
\includegraphics[width=.45\textwidth]{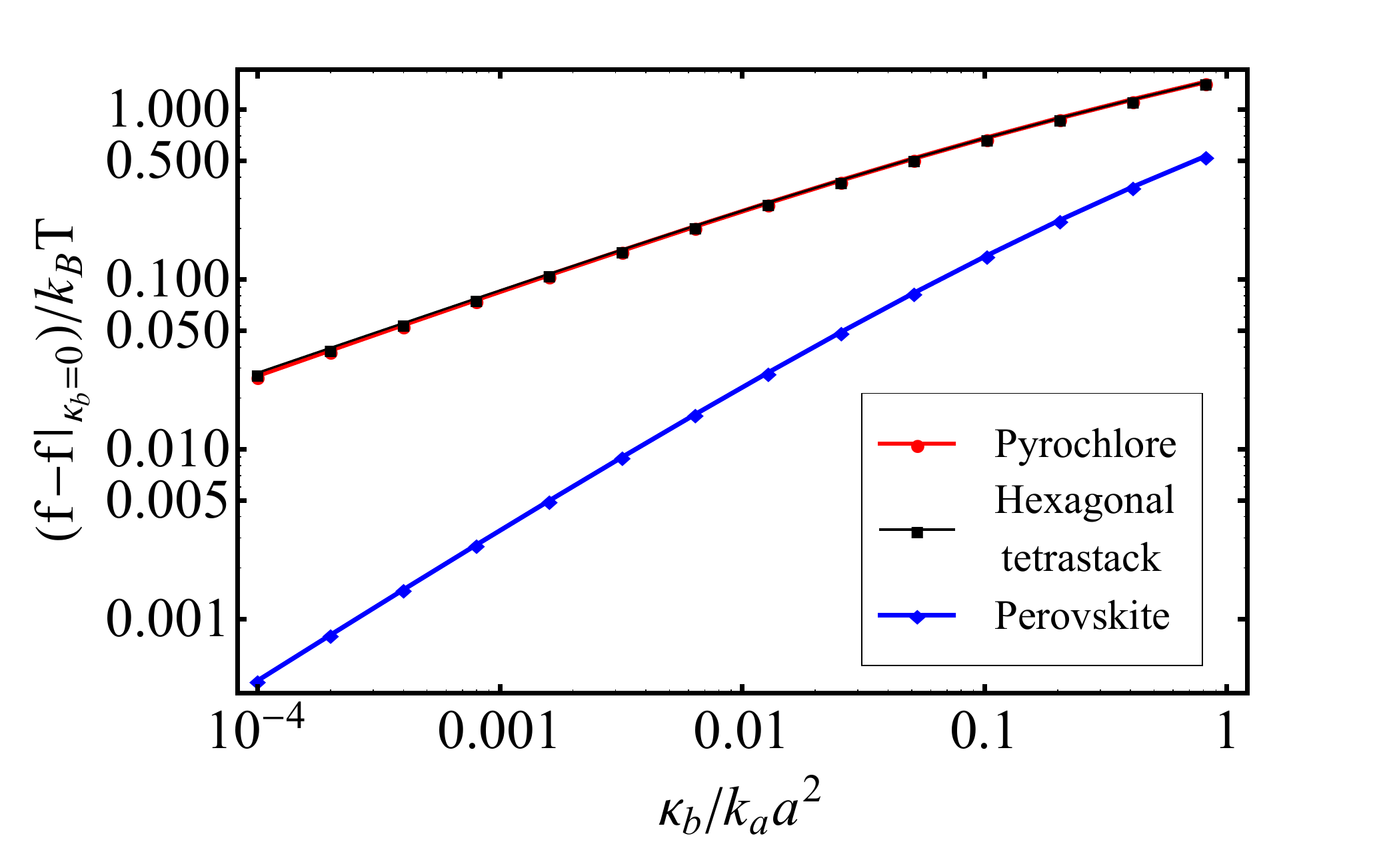}
\caption{
The increase of lattice free energy per particle as a function of the bending stiffness $\kb$. For the pyrochlore (black curve) and hexagonal tetrastack lattices (red curve), there are nearly identical $\sim \sqrt{\kb}$ contributions from surfaces of modes that are zero-energy for $\kb = 0$. For the perovskite lattice (blue curve), which has only a line of such modes, the free energy increase is linear for small $\kb$.
}
\label{fig:fedif}
\end{figure}

We calculated the change of lattice free energy per particle as a function of bending modulus which comes from orientational entropy of the particles and is controlled by the patch size as show in Fig.~\ref{fig:patchsizevsmodulus}, and the results are shown in Fig.~\ref{fig:fedif}.  This reflects the decreased entropy from limiting lattice distortions to ones that keep all bonds within attractive patches. 
For the pyrochlore lattice, this increase is $O(\sqrt{\kb})$ for small $\kb$, reflecting the modifications to the zero modes. The HT lattice, with its similar structure, has nearly identical free energies, suggesting that mixtures of pyrochlore and HT structures may form, as noted earlier by~\cite{Romano2012}.

In contrast to the isostatic pyrochlore and HT lattices, the perovskite lattice with only a number $O(L)$ of such zero modes has a slower increase $O(\kb)$ in the free energy as bending modulus increases.

\subsection{Comparison to close-packed structure}
\begin{figure}
\centering
\subfigure[]{\includegraphics[width=.2\textwidth]{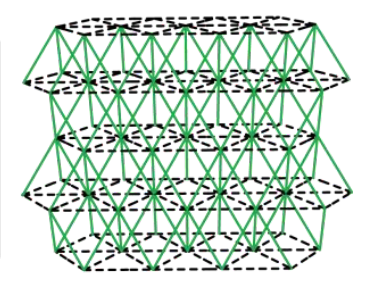}}
\subfigure[]{\includegraphics[width=.2\textwidth]{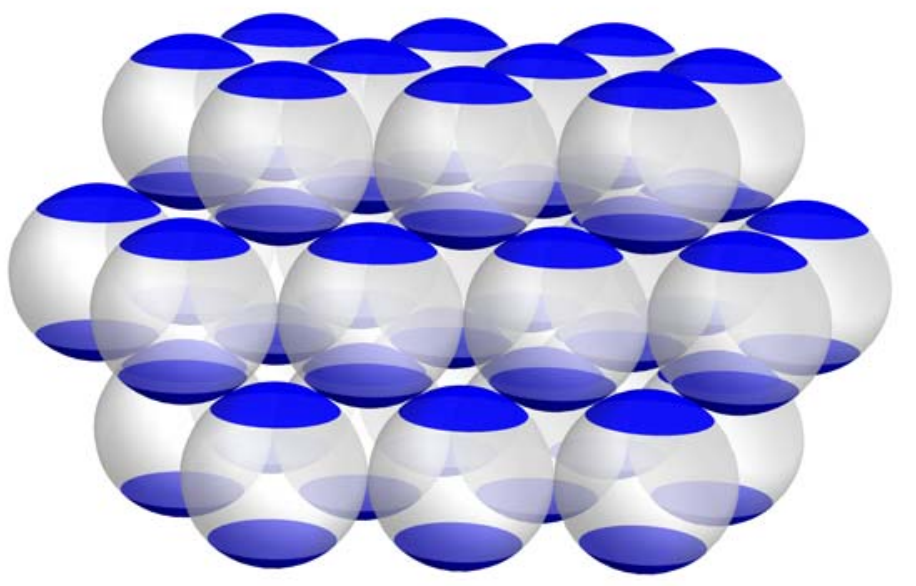}}
\subfigure[]{\includegraphics[width=.2\textwidth]{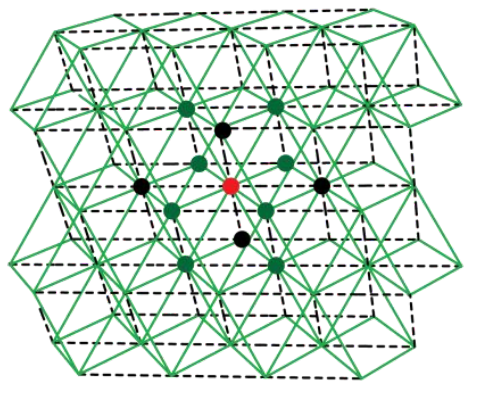}}
\subfigure[]{\includegraphics[width=.2\textwidth]{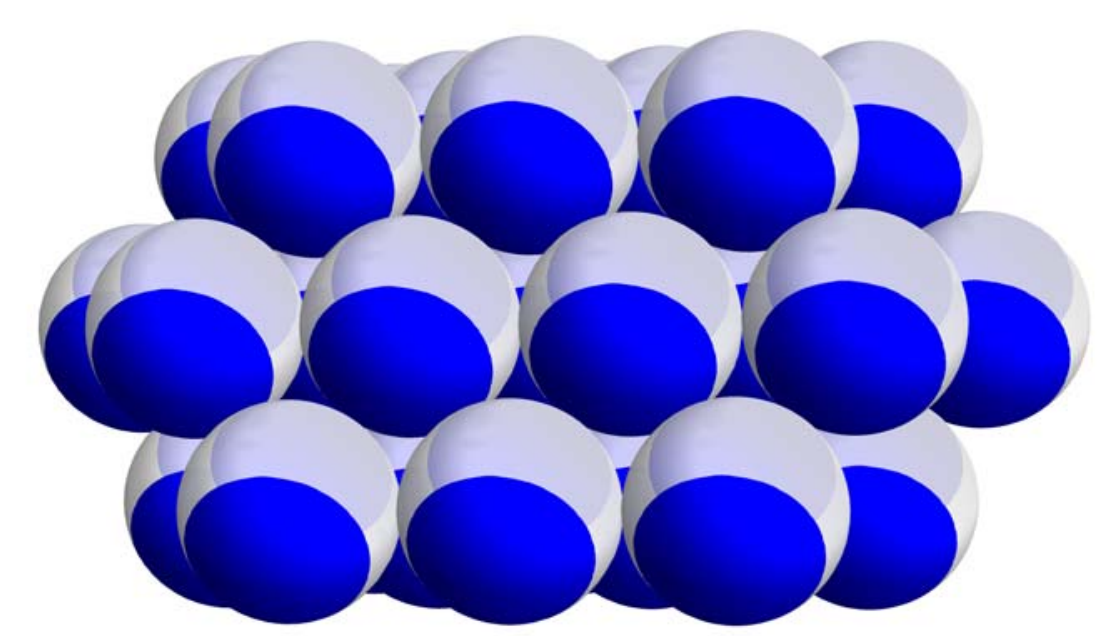}}
\caption{
Close-packed FCC lattices formed by triblock Janus particles. (a) and (b) show the FCC lattice in which each triblock Janus particle has 6 bonds in its attractive patches, corresponding to the patch size that supports the pyrochlore/HT lattices at zero pressure.  (c) and (d) show the FCC lattice in which each triblock Janus particle has 8 bonds in its attractive patches, corresponding to the patch size that supports the perovskite lattices at zero pressure.  In (a) and (c), attractive bonds are denoted by green lines and repulsive bonds by dashed black lines.  The green dots in (b) represent sites attractively bonded to the red site, and the black dots to repulsively bonded sites.
}
\label{fig:fccs}
\end{figure}

\begin{figure}[hh]
\centering
\subfigure{\includegraphics[width=.35\textwidth]{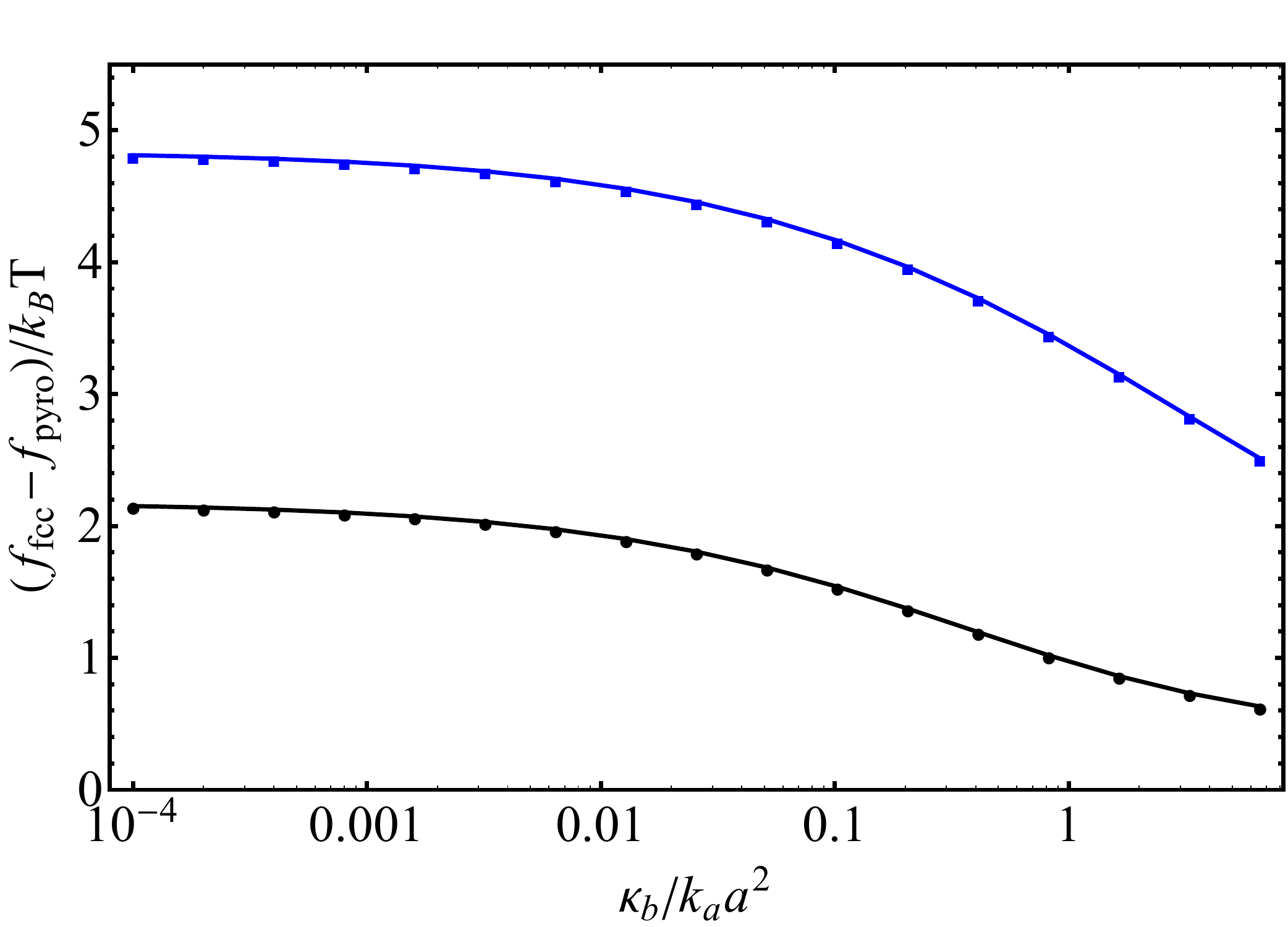}}
\subfigure{\includegraphics[width=.35\textwidth]{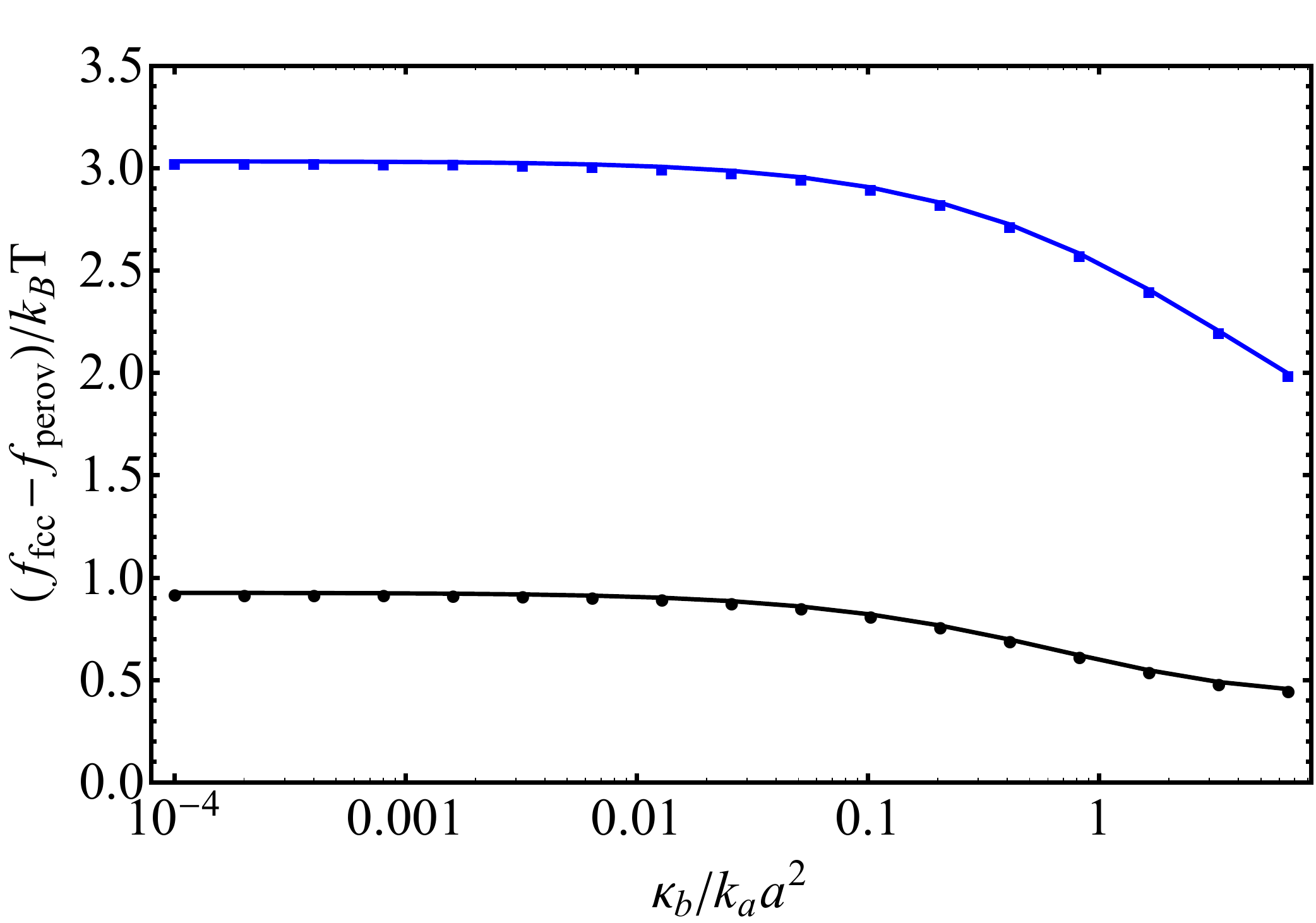}}
\caption{(a) Free energy difference between the FCC and the pyrochlore lattices as a function of $\kb/k_a a^2$ for $k_r = 20$ (top, blue curve) and $k_r = 1$ (bottom, black curve). (b) Free energy difference between the FCC and the perovskite lattices as a function of $\kb/k_a a^2$ for $k_r = 20$ (top, blue curve) and $k_r = 1$ (bottom, black curve).
}
\label{fig:fegainbykb}
\end{figure}

Competing with the open structures above are the close-packed face-centered cubic (FCC) lattice, as well as the hexagonal close-packed lattice, which has nearly identical free energies. In such a lattice, each triblock Janus particle has twelve nearest neighbors, even though not all of them lie within the attractive patches, as can be seen in Fig.~\ref{fig:fccs}. Such a close-packed structure is favored by a finite pressure, but is disfavored by entropic considerations. These additional repulsive particle-particle contacts, which have spring constant $k_r \approx 20 k_a$~\cite{Mao2013}, substantially inhibit the fluctuations in the positions of the triblock Janus particles. This leads to rises in free energy as shown in Fig.~\ref{fig:fegainbykb}. Both the bending modulus and the repulsive bonds tend to limit the position fluctuations of the triblock Janus particles in the same way, so that the bending modulus has a greater effect on the open structures than on the FCC lattice, leading to closer free energies as $\kb$ increases. The free energy differences are greater for the pyrochlore than the perovskite lattices, reflecting the fact that triblock Janus particles with patch size appropriate to the pyrochlore lattice have six repulsive bonds per particle in the FCC lattice, while perovskite-size patches lead to four repulsive bonds in the FCC lattice.

\begin{figure}
\vspace{10pt}
\centering
\includegraphics[width=.45\textwidth]{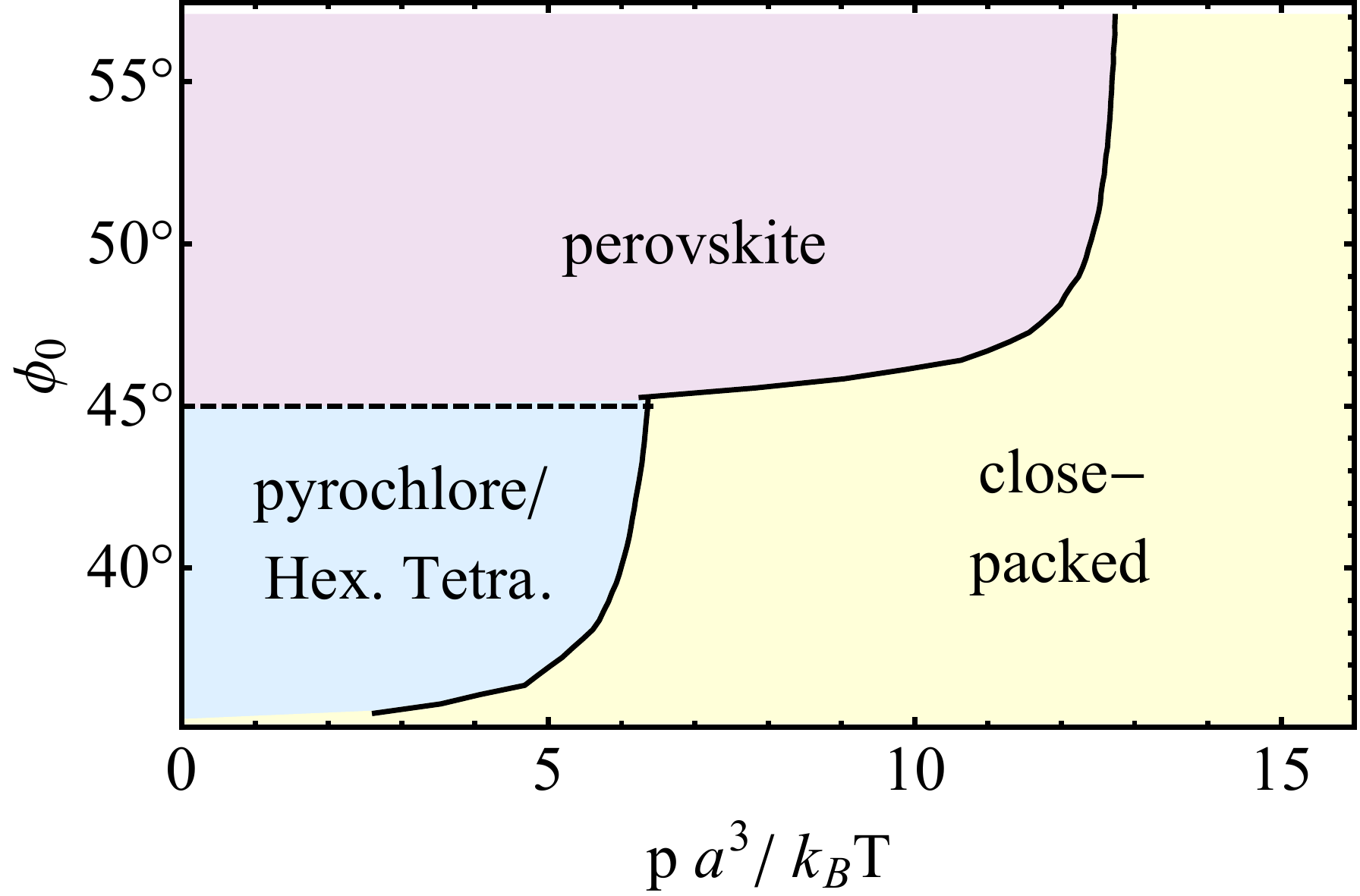}
\caption{
Equilibrium phase diagram showing the transition between open and close-packed lattices. Open lattices are more able to fluctuate and are favored at high temperature, whereas high pressures favor the close-packed face-centered cubic lattice.
}
\label{fig:phasediagram}
\end{figure}

Transforming to a finite-pressure system via a Legendre transformation, one obtains the Gibbs free energy per particle
\begin{eqnarray}
\label{eq:gibbs}
g= f + p v,
\end{eqnarray}
\noindent where $p$ is the pressure and $v$ the volume per particle. From this, one may determine the dominant phase at a given patch size and pressure, as shown in Figure~\ref{fig:phasediagram}.  At low pressures, either the perovskite lattice or a mixture of pyrochlore and HT (depending on whether the patch size is enough to allow four bonds) is present. The perovskite lattice, despite having a lower free energy difference (cf. Fig.~\ref{fig:fegainbykb}) is present at higher pressures than the pyrochlore owing to its denser structure.

Note that unlike the dimensionless moduli against bond length fluctuations, $k a^2/2 \pi k_B T$, which decrease with increasing temperature, the dimensionless bending modulus has entropic rather than energetic origins and so does not scale with temperature. This leads to nontrivial temperature dependence in the dynamical matrix and corrections to the phase diagram in Fig.~\ref{fig:phasediagram}, but because the bending modulus is much weaker than the central-force moduli at all relevant temperatures, this effect on the free energy is very small.

\section{Conclusion}
In this paper, we discuss the self-assembly of three-
dimensional open crystal lattices from patchy colloidal
particles. We have shown that entropy of the particles
not only stabilize these open lattices against mechanical instability, but also favor open structures over close-
packed ones, given that they are energetically equivalent.

The entropic mechanisms discussed in this paper provide a framework for designing simple building blocks for
the self-assembly of colloidal open lattices. We show that
by harnessing entropic effects, open lattices can spontaneously form without having to control surfacing pattern
of colloidal particles beyond currently available experimental techniques: regular open lattices are favored by
entropy.


\begin{thebibliography}{27}
\expandafter\ifx\csname natexlab\endcsname\relax\def\natexlab#1{#1}\fi
\expandafter\ifx\csname bibnamefont\endcsname\relax
  \def\bibnamefont#1{#1}\fi
\expandafter\ifx\csname bibfnamefont\endcsname\relax
  \def\bibfnamefont#1{#1}\fi
\expandafter\ifx\csname citenamefont\endcsname\relax
  \def\citenamefont#1{#1}\fi
\expandafter\ifx\csname url\endcsname\relax
  \def\url#1{\texttt{#1}}\fi
\expandafter\ifx\csname urlprefix\endcsname\relax\def\urlprefix{URL }\fi
\providecommand{\bibinfo}[2]{#2}
\providecommand{\eprint}[2][]{\url{#2}}

\bibitem[{\citenamefont{Greaves et~al.}(2011)\citenamefont{Greaves, Greer,
  Lakes, and Rouxel}}]{Greaves2011}
\bibinfo{author}{\bibfnamefont{G.~N.} \bibnamefont{Greaves}},
  \bibinfo{author}{\bibfnamefont{A.~L.} \bibnamefont{Greer}},
  \bibinfo{author}{\bibfnamefont{R.~S.} \bibnamefont{Lakes}}, \bibnamefont{and}
  \bibinfo{author}{\bibfnamefont{T.}~\bibnamefont{Rouxel}},
  \bibinfo{journal}{Nat. Mater.} \textbf{\bibinfo{volume}{10}},
  \bibinfo{pages}{823} (\bibinfo{year}{2011}).

\bibitem[{\citenamefont{Sun et~al.}(2012)\citenamefont{Sun, Souslov, Mao, and
  Lubensky}}]{Sun2012}
\bibinfo{author}{\bibfnamefont{K.}~\bibnamefont{Sun}},
  \bibinfo{author}{\bibfnamefont{A.}~\bibnamefont{Souslov}},
  \bibinfo{author}{\bibfnamefont{X.}~\bibnamefont{Mao}}, \bibnamefont{and}
  \bibinfo{author}{\bibfnamefont{T.~C.} \bibnamefont{Lubensky}},
  \bibinfo{journal}{Proc. Natl. Acad. Sci. U. S. A.}
  \textbf{\bibinfo{volume}{109}}, \bibinfo{pages}{12369}
  (\bibinfo{year}{2012}).

\bibitem[{\citenamefont{Kapko et~al.}(2009)\citenamefont{Kapko, Treacy, Thorpe,
  and Guest}}]{Kapko2009}
\bibinfo{author}{\bibfnamefont{V.}~\bibnamefont{Kapko}},
  \bibinfo{author}{\bibfnamefont{M.}~\bibnamefont{Treacy}},
  \bibinfo{author}{\bibfnamefont{M.}~\bibnamefont{Thorpe}}, \bibnamefont{and}
  \bibinfo{author}{\bibfnamefont{S.}~\bibnamefont{Guest}},
  \bibinfo{journal}{Proc. R. Soc. London, Ser. A}
  \textbf{\bibinfo{volume}{465}}, \bibinfo{pages}{3517} (\bibinfo{year}{2009}).

\bibitem[{\citenamefont{Grima et~al.}(2007)\citenamefont{Grima, Gatt, Zammit,
  Williams, Evans, Alderson, and Walton}}]{Grima2007}
\bibinfo{author}{\bibfnamefont{J.~N.} \bibnamefont{Grima}},
  \bibinfo{author}{\bibfnamefont{R.}~\bibnamefont{Gatt}},
  \bibinfo{author}{\bibfnamefont{V.}~\bibnamefont{Zammit}},
  \bibinfo{author}{\bibfnamefont{J.~J.} \bibnamefont{Williams}},
  \bibinfo{author}{\bibfnamefont{K.~E.} \bibnamefont{Evans}},
  \bibinfo{author}{\bibfnamefont{A.}~\bibnamefont{Alderson}}, \bibnamefont{and}
  \bibinfo{author}{\bibfnamefont{R.~I.} \bibnamefont{Walton}},
  \bibinfo{journal}{J. Appl. Phys.} \textbf{\bibinfo{volume}{101}}
  (\bibinfo{year}{2007}).

\bibitem[{\citenamefont{Ernst et~al.}(1998)\citenamefont{Ernst, Broholm,
  Kowach, and Ramirez}}]{Ernst1998}
\bibinfo{author}{\bibfnamefont{G.}~\bibnamefont{Ernst}},
  \bibinfo{author}{\bibfnamefont{C.}~\bibnamefont{Broholm}},
  \bibinfo{author}{\bibfnamefont{G.~R.} \bibnamefont{Kowach}},
  \bibnamefont{and} \bibinfo{author}{\bibfnamefont{A.~P.}
  \bibnamefont{Ramirez}}, \bibinfo{journal}{Nature}
  \textbf{\bibinfo{volume}{396}}, \bibinfo{pages}{147} (\bibinfo{year}{1998}).

\bibitem[{\citenamefont{Hammonds et~al.}(1996)\citenamefont{Hammonds, Dove,
  Giddy, Heine, and B.}}]{Hammonds1996}
\bibinfo{author}{\bibfnamefont{K.~D.} \bibnamefont{Hammonds}},
  \bibinfo{author}{\bibfnamefont{M.~T.} \bibnamefont{Dove}},
  \bibinfo{author}{\bibfnamefont{A.~P.} \bibnamefont{Giddy}},
  \bibinfo{author}{\bibfnamefont{V.}~\bibnamefont{Heine}}, \bibnamefont{and}
  \bibinfo{author}{\bibfnamefont{W.}~\bibnamefont{B.}}, \bibinfo{journal}{Am.
  Mineral.} \textbf{\bibinfo{volume}{81}}, \bibinfo{pages}{1057}
  (\bibinfo{year}{1996}).

\bibitem[{\citenamefont{Davis and Lobo}(1992)}]{Davis1992}
\bibinfo{author}{\bibfnamefont{M.~E.} \bibnamefont{Davis}} \bibnamefont{and}
  \bibinfo{author}{\bibfnamefont{R.~F.} \bibnamefont{Lobo}},
  \bibinfo{journal}{Chem. Mater.} \textbf{\bibinfo{volume}{4}},
  \bibinfo{pages}{756} (\bibinfo{year}{1992}).

\bibitem[{\citenamefont{McDowellboyer et~al.}(1986)\citenamefont{McDowellboyer,
  Hunt, and Sitar}}]{McDowellboyer1986}
\bibinfo{author}{\bibfnamefont{L.~M.} \bibnamefont{McDowellboyer}},
  \bibinfo{author}{\bibfnamefont{J.~R.} \bibnamefont{Hunt}}, \bibnamefont{and}
  \bibinfo{author}{\bibfnamefont{N.}~\bibnamefont{Sitar}},
  \bibinfo{journal}{Water Resour. Res.} \textbf{\bibinfo{volume}{22}},
  \bibinfo{pages}{1901} (\bibinfo{year}{1986}).

\bibitem[{\citenamefont{Breck}(1973)}]{Breck1973}
\bibinfo{author}{\bibfnamefont{D.~W.} \bibnamefont{Breck}},
  \emph{\bibinfo{title}{Zeolite molecular sieves: structure, chemistry, and
  use}}, vol.~\bibinfo{volume}{4} (\bibinfo{publisher}{Wiley New York},
  \bibinfo{year}{1973}).

\bibitem[{\citenamefont{Joannopoulos et~al.}(1997)\citenamefont{Joannopoulos,
  Villeneuve, and Fan}}]{Joannopoulos1997}
\bibinfo{author}{\bibfnamefont{J.~D.} \bibnamefont{Joannopoulos}},
  \bibinfo{author}{\bibfnamefont{P.~R.} \bibnamefont{Villeneuve}},
  \bibnamefont{and} \bibinfo{author}{\bibfnamefont{S.~H.} \bibnamefont{Fan}},
  \bibinfo{journal}{Nature} \textbf{\bibinfo{volume}{386}},
  \bibinfo{pages}{143} (\bibinfo{year}{1997}).

\bibitem[{\citenamefont{Moroz}(2002)}]{Moroz2002}
\bibinfo{author}{\bibfnamefont{A.}~\bibnamefont{Moroz}},
  \bibinfo{journal}{Phys. Rev. B} \textbf{\bibinfo{volume}{66}}
  (\bibinfo{year}{2002}).

\bibitem[{\citenamefont{Garcia-Adeva}(2006)}]{Garcia-Adeva2006}
\bibinfo{author}{\bibfnamefont{A.~J.} \bibnamefont{Garcia-Adeva}},
  \bibinfo{journal}{New J. Phys.} \textbf{\bibinfo{volume}{8}}
  (\bibinfo{year}{2006}).

\bibitem[{\citenamefont{Galisteo-Lopez
  et~al.}(2011)\citenamefont{Galisteo-Lopez, Ibisate, Sapienza, Froufe-Perez,
  Blanco, and Lopez}}]{Galisteo-Lopez2011}
\bibinfo{author}{\bibfnamefont{J.~F.} \bibnamefont{Galisteo-Lopez}},
  \bibinfo{author}{\bibfnamefont{M.}~\bibnamefont{Ibisate}},
  \bibinfo{author}{\bibfnamefont{R.}~\bibnamefont{Sapienza}},
  \bibinfo{author}{\bibfnamefont{L.~S.} \bibnamefont{Froufe-Perez}},
  \bibinfo{author}{\bibfnamefont{A.}~\bibnamefont{Blanco}}, \bibnamefont{and}
  \bibinfo{author}{\bibfnamefont{C.}~\bibnamefont{Lopez}},
  \bibinfo{journal}{Adv. Mater.} \textbf{\bibinfo{volume}{23}},
  \bibinfo{pages}{30} (\bibinfo{year}{2011}).

\bibitem[{\citenamefont{Maxwell}(1864)}]{Maxwell1864}
\bibinfo{author}{\bibfnamefont{J.~C.} \bibnamefont{Maxwell}},
  \bibinfo{journal}{Philos. Mag.} \textbf{\bibinfo{volume}{27}},
  \bibinfo{pages}{294} (\bibinfo{year}{1864}).

\bibitem[{\citenamefont{Cohn and Kumar}(2009)}]{Cohn2009}
\bibinfo{author}{\bibfnamefont{H.}~\bibnamefont{Cohn}} \bibnamefont{and}
  \bibinfo{author}{\bibfnamefont{A.}~\bibnamefont{Kumar}},
  \bibinfo{journal}{Proc. Natl. Acad. Sci. U.S.A.}
  \textbf{\bibinfo{volume}{106}}, \bibinfo{pages}{9570} (\bibinfo{year}{2009}).

\bibitem[{\citenamefont{Edlund et~al.}(2011)\citenamefont{Edlund, Lindgren, and
  Jacobi}}]{Edlund2011}
\bibinfo{author}{\bibfnamefont{E.}~\bibnamefont{Edlund}},
  \bibinfo{author}{\bibfnamefont{O.}~\bibnamefont{Lindgren}}, \bibnamefont{and}
  \bibinfo{author}{\bibfnamefont{M.~N.} \bibnamefont{Jacobi}},
  \bibinfo{journal}{Phys. Rev. Lett.} \textbf{\bibinfo{volume}{107}},
  \bibinfo{pages}{085501} (\bibinfo{year}{2011}).

\bibitem[{\citenamefont{Torquato}(2009)}]{Torquato2009}
\bibinfo{author}{\bibfnamefont{S.}~\bibnamefont{Torquato}},
  \bibinfo{journal}{Soft Matter} \textbf{\bibinfo{volume}{5}},
  \bibinfo{pages}{1157} (\bibinfo{year}{2009}).

\bibitem[{\citenamefont{Velikov et~al.}(2002)\citenamefont{Velikov, Christova,
  Dullens, and van Blaaderen}}]{Velikov2002}
\bibinfo{author}{\bibfnamefont{K.~P.} \bibnamefont{Velikov}},
  \bibinfo{author}{\bibfnamefont{C.~G.} \bibnamefont{Christova}},
  \bibinfo{author}{\bibfnamefont{R.~P.~A.} \bibnamefont{Dullens}},
  \bibnamefont{and} \bibinfo{author}{\bibfnamefont{A.}~\bibnamefont{van
  Blaaderen}}, \bibinfo{journal}{Science} \textbf{\bibinfo{volume}{296}},
  \bibinfo{pages}{106} (\bibinfo{year}{2002}).

\bibitem[{\citenamefont{Glotzer and Solomon}(2007)}]{Glotzer2007}
\bibinfo{author}{\bibfnamefont{S.~C.} \bibnamefont{Glotzer}} \bibnamefont{and}
  \bibinfo{author}{\bibfnamefont{M.~J.} \bibnamefont{Solomon}},
  \bibinfo{journal}{Nat. Mater.} \textbf{\bibinfo{volume}{6}},
  \bibinfo{pages}{557} (\bibinfo{year}{2007}).

\bibitem[{\citenamefont{Romano and Sciortino}(2012)}]{Romano2012}
\bibinfo{author}{\bibfnamefont{F.}~\bibnamefont{Romano}} \bibnamefont{and}
  \bibinfo{author}{\bibfnamefont{F.}~\bibnamefont{Sciortino}},
  \bibinfo{journal}{Nat. Commun.} \textbf{\bibinfo{volume}{3}}
  (\bibinfo{year}{2012}).

\bibitem[{\citenamefont{Tkachenko}(2002)}]{Tkachenko2002}
\bibinfo{author}{\bibfnamefont{A.~V.} \bibnamefont{Tkachenko}},
  \bibinfo{journal}{Phys. Rev. Lett.} \textbf{\bibinfo{volume}{89}},
  \bibinfo{pages}{148303} (\bibinfo{year}{2002}).

\bibitem[{\citenamefont{Chen et~al.}(2011{\natexlab{a}})\citenamefont{Chen,
  Bae, and Granick}}]{Chen2011}
\bibinfo{author}{\bibfnamefont{Q.}~\bibnamefont{Chen}},
  \bibinfo{author}{\bibfnamefont{S.}~\bibnamefont{Bae}}, \bibnamefont{and}
  \bibinfo{author}{\bibfnamefont{S.}~\bibnamefont{Granick}},
  \bibinfo{journal}{Nature} \textbf{\bibinfo{volume}{469}},
  \bibinfo{pages}{381} (\bibinfo{year}{2011}{\natexlab{a}}).

\bibitem[{\citenamefont{Jiang et~al.}(2010)\citenamefont{Jiang, Chen, Tripathy,
  Luijten, Schweizer, and Granick}}]{Jiang2010}
\bibinfo{author}{\bibfnamefont{S.}~\bibnamefont{Jiang}},
  \bibinfo{author}{\bibfnamefont{Q.}~\bibnamefont{Chen}},
  \bibinfo{author}{\bibfnamefont{M.}~\bibnamefont{Tripathy}},
  \bibinfo{author}{\bibfnamefont{E.}~\bibnamefont{Luijten}},
  \bibinfo{author}{\bibfnamefont{K.~S.} \bibnamefont{Schweizer}},
  \bibnamefont{and} \bibinfo{author}{\bibfnamefont{S.}~\bibnamefont{Granick}},
  \bibinfo{journal}{Advanced Materials} \textbf{\bibinfo{volume}{22}},
  \bibinfo{pages}{1060} (\bibinfo{year}{2010}).

\bibitem[{\citenamefont{Chen et~al.}(2011{\natexlab{b}})\citenamefont{Chen,
  Diesel, Whitmer, Bae, Luijten, and Granick}}]{Chen2011b}
\bibinfo{author}{\bibfnamefont{Q.}~\bibnamefont{Chen}},
  \bibinfo{author}{\bibfnamefont{E.}~\bibnamefont{Diesel}},
  \bibinfo{author}{\bibfnamefont{J.~K.} \bibnamefont{Whitmer}},
  \bibinfo{author}{\bibfnamefont{S.~C.} \bibnamefont{Bae}},
  \bibinfo{author}{\bibfnamefont{E.}~\bibnamefont{Luijten}}, \bibnamefont{and}
  \bibinfo{author}{\bibfnamefont{S.}~\bibnamefont{Granick}},
  \bibinfo{journal}{J. Am. Chem. Soc.} \textbf{\bibinfo{volume}{133}},
  \bibinfo{pages}{7725} (\bibinfo{year}{2011}{\natexlab{b}}).

\bibitem[{\citenamefont{Mao et~al.}(2013)\citenamefont{Mao, Chen, and
  Granick}}]{Mao2013}
\bibinfo{author}{\bibfnamefont{X.}~\bibnamefont{Mao}},
  \bibinfo{author}{\bibfnamefont{Q.}~\bibnamefont{Chen}}, \bibnamefont{and}
  \bibinfo{author}{\bibfnamefont{S.}~\bibnamefont{Granick}},
  \bibinfo{journal}{Nature Materials} \textbf{\bibinfo{volume}{7}},
  \bibinfo{pages}{217} (\bibinfo{year}{2013}).

\bibitem[{\citenamefont{Mao}(2013)}]{Mao2013b}
\bibinfo{author}{\bibfnamefont{X.}~\bibnamefont{Mao}}, \bibinfo{journal}{Phys.
  Rev. E} \textbf{\bibinfo{volume}{87}}, \bibinfo{pages}{062319}
  (\bibinfo{year}{2013}).

\bibitem[{\citenamefont{Ashcroft and Mermin}(1976)}]{Ashcroft1976}
\bibinfo{author}{\bibfnamefont{N.~W.} \bibnamefont{Ashcroft}} \bibnamefont{and}
  \bibinfo{author}{\bibfnamefont{N.~D.} \bibnamefont{Mermin}},
  \emph{\bibinfo{title}{Solid State Physics}} (\bibinfo{publisher}{Saunders
  College, Philadelphia}, \bibinfo{year}{1976}), \bibinfo{edition}{1st} ed.

\end{thebibliography}
\end{document}